\RequirePackage{color}
\pdfoutput=1
\documentclass[12pt]{article}
\usepackage{graphicx}
\graphicspath{ {./Figures/} }
\title{Pressure Dependence of Carbon Foam Bonding Strength using Reactive Film}
\date{}
\author{M. Chertok\thanks{chertok@ucdavis.edu}, J. Dioquino, J. Hansen, M. Irving, C. Neher, \and M. Tripathi, Y. Yao, and G. Zheng}

\usepackage{palatino}

\begin{document}

\maketitle
\vspace{-0.2in}
\centering{University of California, Davis\\
One Shields Ave., Davis, CA 95616, USA \\
\vspace{0.2in}
July 3, 2018}
\abstract{Reactive bonding film is a relatively new method of fusing materials with the potential to meet needs of particle tracker mechanics under development due to its resulting tensile strength, thermal conductivity, radiation tolerance, and low mass.  Employing a new apparatus to vary pressure applied to samples during bonding, we find improved ultimate tensile strengths compared with previous results.

\section{Introduction}
	At CERN, replacements of the ATLAS and CMS silicon tracking detectors are required for the upcoming High Luminosity run of the LHC.  Structural components with high thermal conductivity, tensile strength, and radiation tolerance are under study for various detector mechanics aspects.  Carbon foam is a key structural material for these purposes.  In previous studies, we have created and tested a wide variety of carbon foam samples with novel materials such as loaded epoxies and reactive bonding film \cite{thermalpaper,RBFpaper}.
    
	Reactive bonding film (RBF) \cite{RBFpaper} consists of multiple nanolayers of aluminum and nickel, and is activated by an electric spark, generating very high localized temperatures within milliseconds.  Our work demonstrates this heat can be used to melt aluminum foil layered with the RBF, and solder together metals or extrude into a porous medium, such as carbon foam, creating a "lock-and-key" configuration.  These shapes hold the carbon foam layers together mechanically and provide for good thermal conduction.  Radiation tolerance, while not yet tested, is expected to be excellent due to the materials involved.
    
	Our previous trials bonding carbon foam with RBF failed mechanically at low tension because the clamps used to compress the sample during bonding did not apply adequate pressure, preventing the aluminum from extruding deeply  into the foam \cite{RBFpaper}. There were also likely issues with uniformity of applied pressure, but this parameter was not studied independently of pressure magnitude.  Successful samples with good tensile strength resulted from maximum pressure and uniformity.  In this updated study, we constructed an apparatus that allows for much higher applied pressure while bonding, as well as easy and quantifiable adjustment of this pressure. This paper presents results using this bonding apparatus.

\section{Method}
%\subsection{Bonding Procedure}

Our samples are created using an apparatus that applies pressure over the stack of materials to be bonded: one square of Allcomp \cite{allcomp} carbon foam (4 mm thick), two sheets of aluminum foil (each 25 $\mu{\rm m}$ thick), two sheets of Nanofoil RBF from Indium Corporation \cite{indium} (each 60 $\mu{\rm m}$ thick), two sheets of aluminum foil, and another layer of foam, as shown in Figure ~\ref{fig:stack}. Through trial and error with our previous experiments detailed in \cite{RBFpaper}, this configuration was determined to provide the strongest and most consistent bond.

\begin{figure}[!ht]
\centering
    \includegraphics[width=0.6\textwidth]{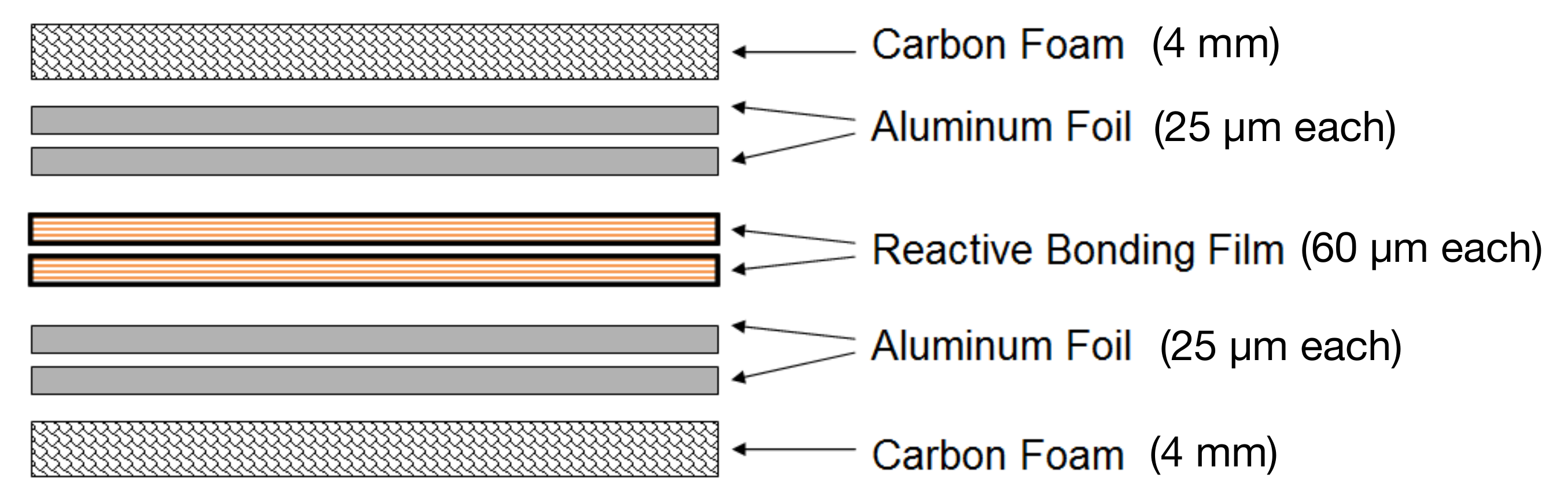}
  \caption{Vertical arrangement of stack to be bonded. Not to scale.}
  \label{fig:stack}
\end{figure}

We vary the pressure applied to the samples during bonding by placing two to ten 11.3-kilogram blocks on the weight table of the apparatus prior to activation. The apparatus is shown in Figure ~\ref{fig:bonding}.  As in \cite{RBFpaper}, the RBF is activated with a 9V battery and the sample is allowed to cool to room temperature before testing.

\begin{figure}[!ht]
\centering
    \includegraphics[width=0.6\textwidth,angle=-90]{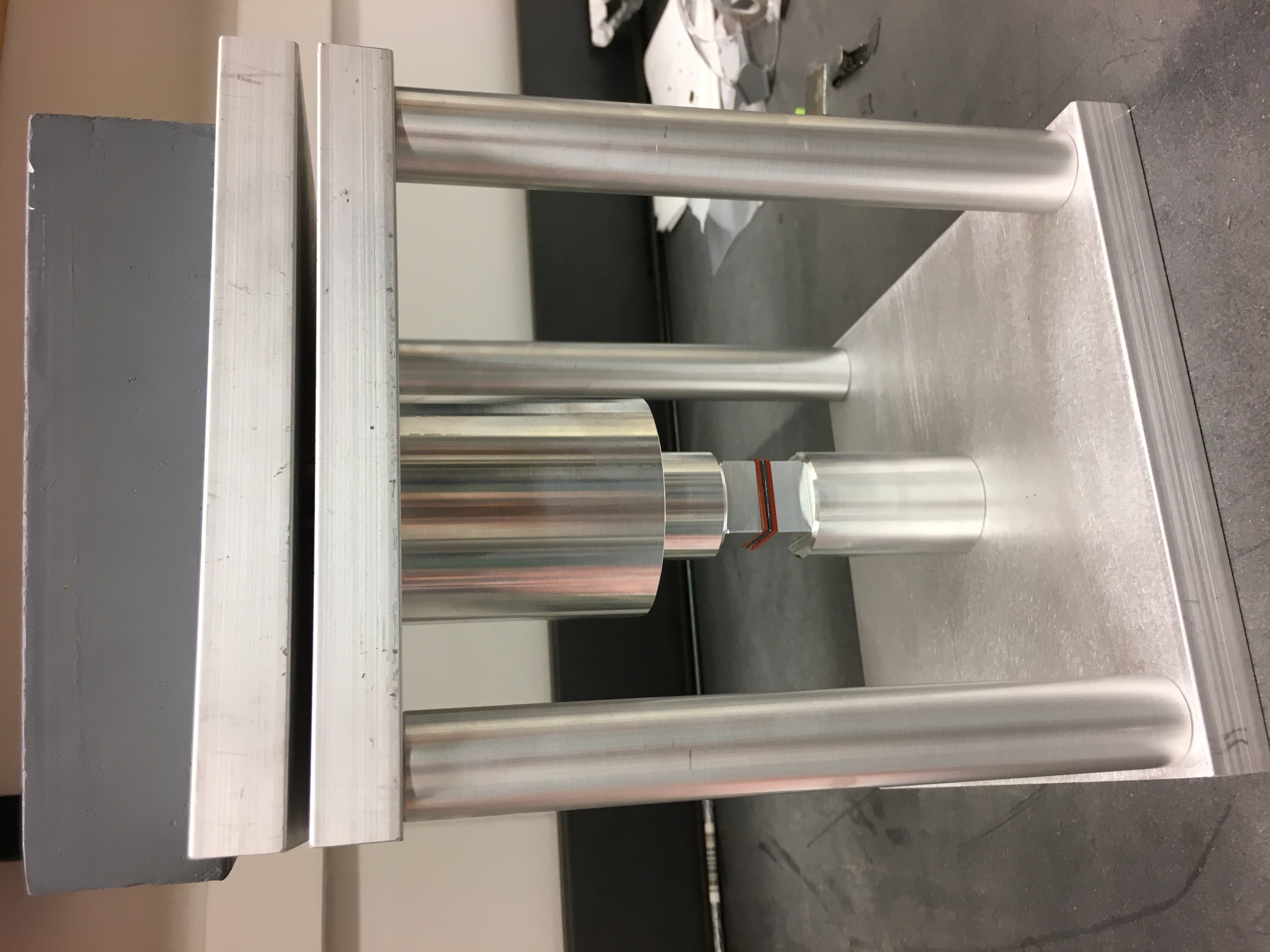}
  \caption{Bonding apparatus with one 11.3-kg block on the weight table. The stack to be bonded is placed between the two protruding 25.4 x 25.4 mm platforms at the center of the apparatus.}
\label{fig:bonding}
\end{figure}

%\subsection{Testing procedure}
For tensile strength testing, we adhere both sides of the sample to aluminum blocks attached to steel hooks, and a pneumatic apparatus pulls the sample in opposite directions until breakage occurs. The device measures the pressure at which the mechanical bond fails with a sensor readout. Calibration of this device is performed using known weights. The tensile tester and calibration curve are shown in Figure ~\ref{fig:tensile}.

\begin{figure}[!ht]
\centering
  \includegraphics[width=0.35\textwidth]{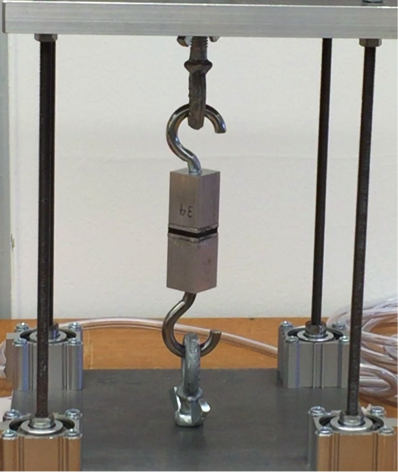}
    \includegraphics[width=0.6\textwidth]{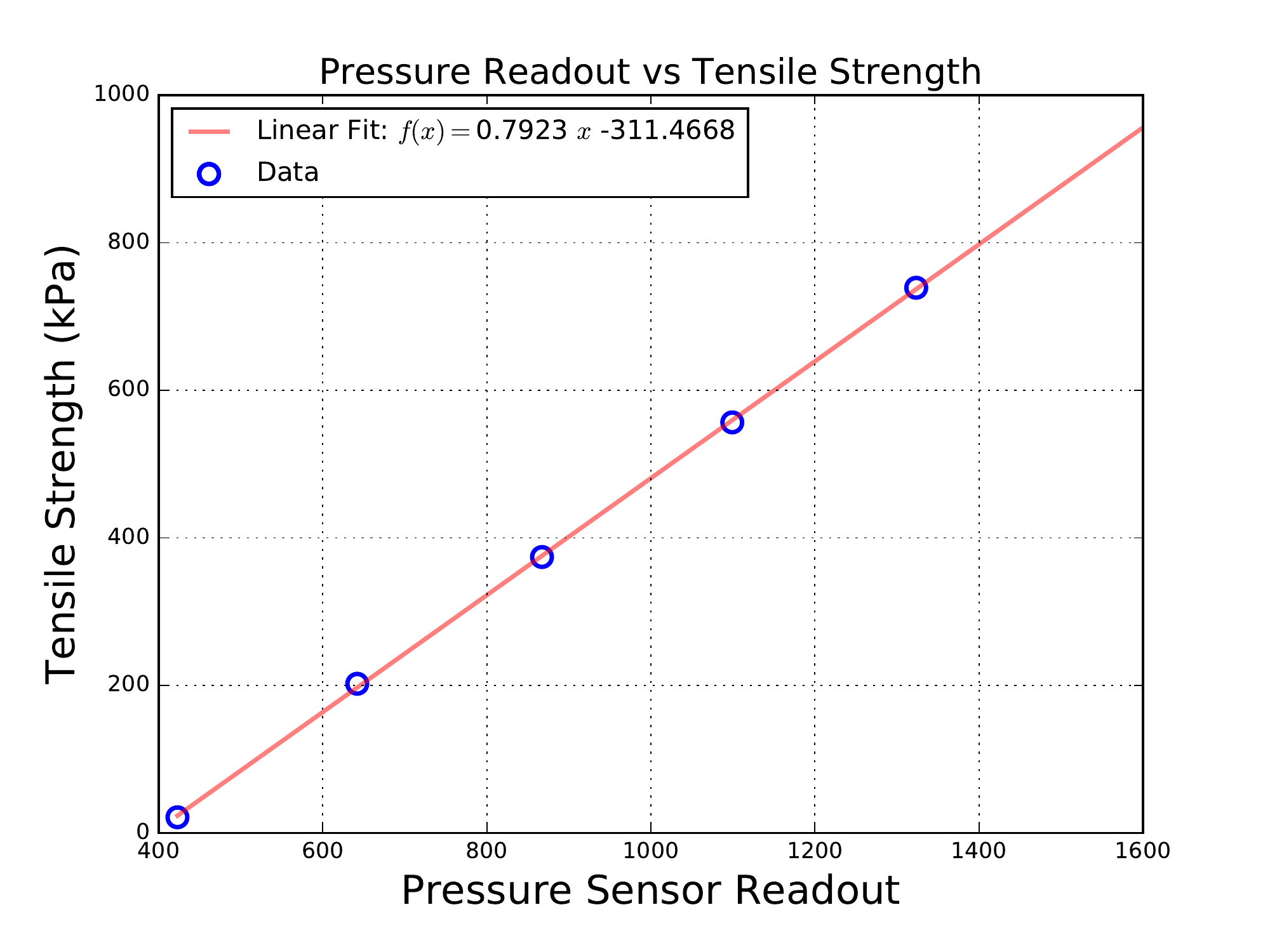}
  \caption{Left: Tensile tester with sample loaded and ready to be tested. Right: Calibration results.}
  \label{fig:tensile}
\end{figure}

%\begin{figure}[!ht]
% \centering
%   \includegraphics[width=0.5\textwidth]{figure4.jpg}
%   \caption{Mechanically-separated stack of carbon foam (top and bottom layers) and aluminum foil plus RBF (center) after bonding.  The center piece is composed of 2 sheets of aluminum foil, 2 sheets of RBF, and another 2 sheets of aluminum foil all fused together by the RBF.  The rough texture of the aluminum foil is caused by hundreds of branching extrusions which locked into the carbon foam.}
%   \label{fig:separated}
% \end{figure}

\section{Results and Discussion}
From tests conducted with 2, 4, 6, 8, and 10 blocks on the weight table during bonding, we find eight blocks (11.3 kg each) results in the maximum ultimate tensile strength (UTS) for a single sample. Table~\ref{tab:results} provides results for each configuration and Figure~\ref{fig:plot} shows a distribution of UTS results for the eleven successful samples.

\begin{table}[!t]
\centering
%\begin{tabular}{@{}|c|c|c|c|c|c|c|@{}}
\begin{tabular}{|c|c|c|c|c|}
\hline 
Sample & \# of blocks applied & Corresponding  & UTS (kPa) \\
       & during bonding & pressure (kPa) & \\

\hline\hline
01 & 2 & 344 &49 \\\hline
02 & 4 & 687 &121 \\\hline
03 & 4 & 687 &167 \\\hline
04 & 4 & 687 &141 \\\hline
05 & 6 & 1031 &124  \\\hline
06 & 6 & 1031 &240  \\\hline
07 & 6 & 1031 &242 \\\hline
08 & 8 & 1375 &200 \\\hline
09 & 8 & 1375 &280 \\\hline
10 & 8 & 1375 &204 \\\hline
11 & 10 & 1718 &162 \\\hline
12 & 10 & 1718 &150 \\\hline

\end{tabular}
\caption{Ultimate tensile strength (UTS) for different numbers of blocks placed on the weight table during bonding. Sample 05 exhibited a fissure across the foam, which likely caused shear failure.  Failure mode analysis showed the foam itself had an unusual structural weakness, so it is not included in Figure~\ref{fig:plot}.}
\label{tab:results}
\end{table}

\begin{figure}[!ht]
\centering
  \includegraphics[width=0.8\textwidth]{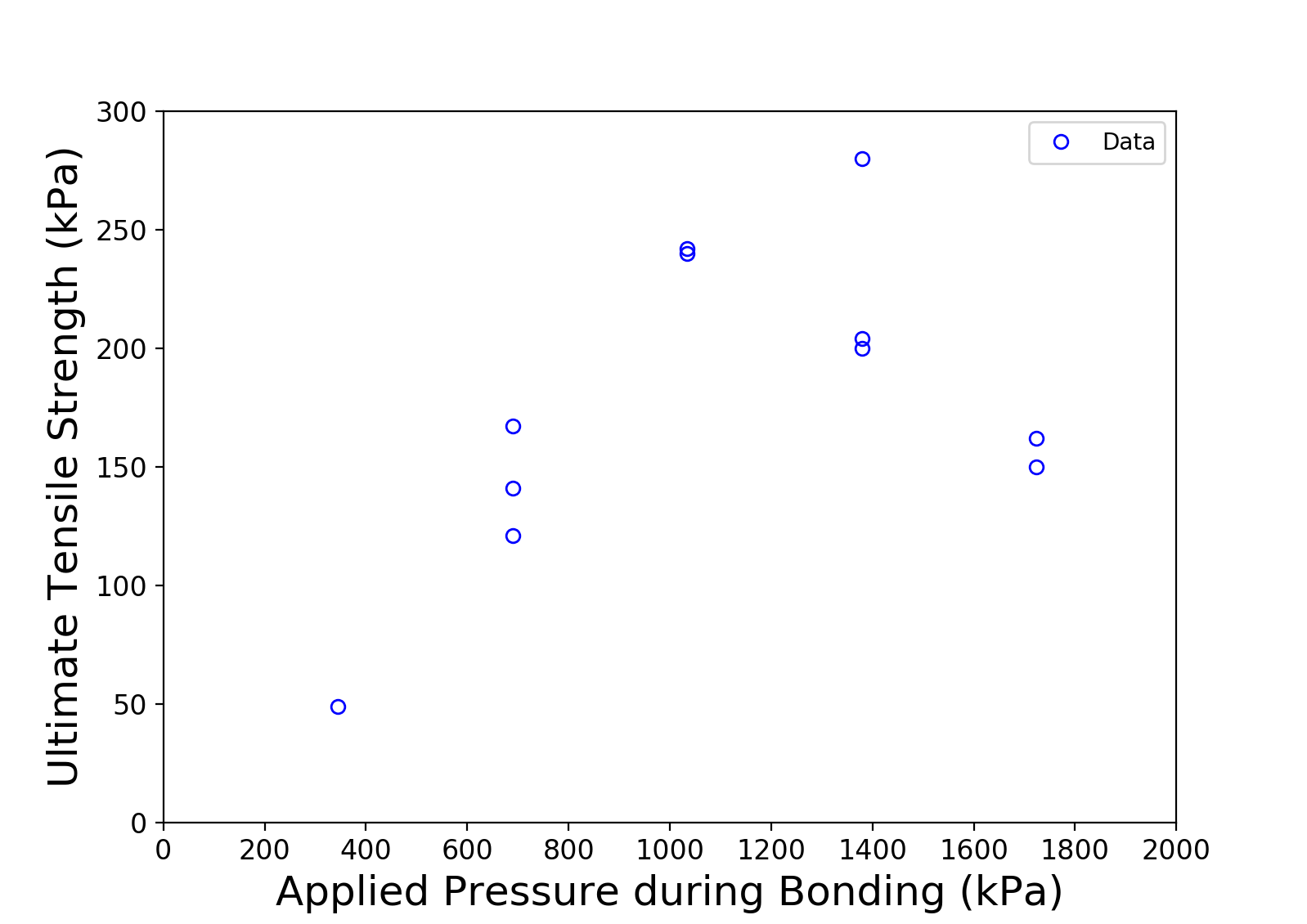}
  \caption{UTS versus pressure applied to weight table for all successful samples.}
  \label{fig:plot}
\end{figure}

Ultimate tensile strength of RBF-fused samples improves substantially with increasing applied pressure up to a point, with 1375 kPa, or eight blocks, being the most effective. We can infer from these results that too little pressure allows the rapid thermal expansion of air to separate the materials as the reaction proceeds\footnote{A slight jump is observed during the activation in this case, corroborating this inference.}, while too much pressure causes the carbon foam to be partially crushed under the weight, likely diminishing the size of the surface pores and limiting sufficient extrusion of aluminum into the foam.  Systematic uncertainties include the precision of the tensile tester pressure readout and its calibration, resulting in a $<5\%$ uncertainty in the measured UTS values.  

These outcomes are consistently stronger mechanically than the samples created and studied previously; in that experiment, the maximum tensile strength achieved was 147 kPa \cite{RBFpaper}. Our new setup solves the problems of magnitude and uniformity of pressure reported earlier, because the weight of the lead blocks is significantly higher than the force applied by clamps, and the precise square platform where the stack is placed ensures that that weight is distributed uniformly across the whole sample. This optimizes the aluminum extrusion process, which provides the mechanical strength.

\section{Conclusions and Future Work}
We have improved the process of bonding carbon foam using RBF by employing an apparatus to apply high and uniform pressure over the sample during bonding.  The optimized bonding procedure consistently produces samples able to withstand $>$200 kPa of tensile pressure before failure. 

Further studies may measure the radiation tolerance for RBF samples and employ similar techniques to bond other materials such as metals and carbon fiber.

\section{Acknowledgments}
RBF materials were provided as part of a research agreement with Indium Corporation.  This work at the University of California, Davis, was supported by U.S. CMS R\&D funds via Fermilab.

%%%%%%%%%%%%%%%%%%%%%%%%%%%%%%%
%\bibliography{References}{}
\bibliographystyle{unsrt}

\end{document}